\documentclass[twocolumn,aps,prl,showpacs]{revtex4-1}

\usepackage{amsfonts}
\usepackage{amsmath}
\usepackage{graphicx}
\usepackage{bm}
\usepackage{amssymb}
\usepackage{dcolumn}
\usepackage{color}
\begin{document}

\title{Sign change of the spin Hall effect due to electron correlation in nonmagnetic CuIr alloys}

\author{Zhuo Xu$^{1,2}$,
Bo Gu$^{1,2}$, Michiyasu Mori$^{1,2}$, Timothy Ziman$^{3,4}$, and Sadamichi Maekawa$^{1,2}$}

\affiliation{$^1$Advanced Science Research Center, Japan Atomic Energy Agency, Tokai 319-1195, Japan
\\ $^{2}$CREST, Japan Science and Technology Agency, Sanbancho, Tokyo 102-0075, Japan
\\ $^{3}$Institut Laue Langevin, Bo\^{\i}te Postale 156, F-38042 Grenobel Cedex 9, France
\\ $^{4}$LPMMC (UMR 5493), Universit\'{e} Grenoble 1 and CNRS, 38042 Grenoble, France
}

\date{\today}
\begin{abstract}
Recently  a {\it positive} spin Hall angle (SHA) of 0.021,
was observed experimentally  in nonmagnetic CuIr alloys [Niimi et al., Phys. Rev. Lett. {\bf 106}, 126601 (2011)]  and attributed predominantly to an  extrinsic skew scattering mechanism,
while a {\it negative} SHA was obtained from  {\it ab initio} calculations [Fedorov {\it et al}., Phys. Rev. B {\bf 88}, 085116 (2013)],
using  consistent definitions of the SHA.
We reconsider the SHA in CuIr alloys,
with  the effects of the local electron correlation $U$ in $5d$ orbitals of Ir impurities,
included by the quantum Monte Carlo method.
We found that the SHA is negative if we ignore such local  electron correlation,
but  becomes positive once $U$ approaches a realistic value.
This may open up a way to control the sign of the SHA by manipulating the occupation number of impurities.
\end{abstract}

\maketitle
The spin Hall effect, which converts charge
current into spin current via the spin-orbit interaction (SOI), is one
of the key phenomena for the further development of
spintronics devices \cite{Dyakonov, Hirsch}.
The spin Hall angle (SHA) describes the conversion efficiency
from the injected longitudinal charge current into the scattered transverse spin current.
Recently  a {\it positive} SHA
of 2.1\% was measured in nonmagnetic CuIr alloys \cite{Niimi-CuIr} and argued to be  predominantly due to  extrinsic skew scattering,
while  {\it negative} SHA of -0.035 and -0.029 were calculated for CuIr alloys,
from the Boltzmann equation and the Kubo-Streda formula respectively \cite{Fedorov},
with consistent definitions \cite{Gu-CuBi,footnote}.
The spin Hall effect in CuIr alloys in experiment is mainly due to  skew scattering \cite{Niimi-CuIr},
which is well described by the phase-shift model of Fert and Levy \cite{Fert-Levy,Fert}.
According to this model, the SHA is proportional to the phase shift parameter $\delta_{1}$,
which was originally taken to be $|\delta_{1}|\simeq$ 0.1 \cite{Fert-Levy, Fert}.
It is given as $\delta_{1} = \pi(N_{p}^{Ir}-N_{p}^{Cu})/6$, where $N_{p}^{Ir}$ and $N_{p}^{Cu}$ are the occupation numbers for the
$6p$ orbitals around an  Ir impurity and the $4p$ orbitals of Cu host, respectively. In the atomic limit $N_{p}^{Ir}=N_{p}^{Cu}=0$,
while in the CuIr alloys $N_{p}^{Ir}$ and $N_{p}^{Cu}$ become finite, but small, due to the  mixing with the other orbitals.
To predict the sign of the SHA in CuIr alloys, we therefore need to calculate precisely the sign of $\delta_1$ \cite{Fert-Levy, Fert}.

\begin{figure}[tbp]
\includegraphics[width = 5.5 cm]{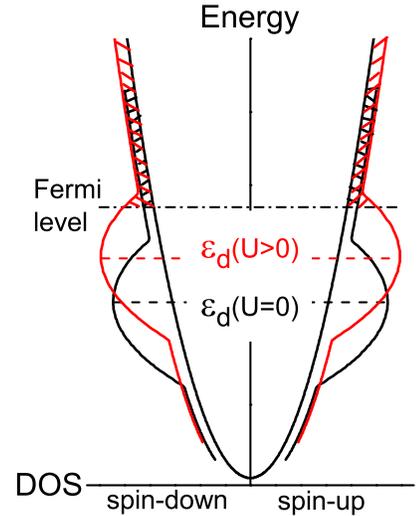}
\caption{(color online).
Schematic picture of the density of states (DOS) of spin-up and spin-down electrons,
in the non-magnetic CuIr alloys. The DOS of Ir $5d$ orbitals are shown by the outer curves
 for the on-site Coulomb repulsion $U$=0 (black) and $U>$0 (red), respectively.
 $\epsilon_d$ is the impurity level in the Hartree-Fock approximation.
The occupation number of Ir $5d$ decreases as the correlation $U$ increases.}
\label{F-schematic}
\end{figure}

In this paper, we argue that the question of the {\it sign} of
the phase shift parameter $\delta_1$ and the SHA in CuIr alloys is highly non-trivial.
Our idea is shown schematically in Fig. {\ref{F-schematic}}.
The nonmagnetic CuIr alloys can be described by an Anderson impurity model \cite{Anderson}, where Ir is an impurity.
In a simplified Anderson model within the Hartree-Fock approximation, it is clear that
due to the on-site Coulomb correlation $U$ the impurity level increases, while the impurity occupation number decreases \cite{Anderson}.

For Ir impurities in Cu, we define the occupation numbers of the $5d$ ($N_{d}^{Ir}$), $6s$ ($N_{s}^{Ir}$) and $6p$ ($N_{p}^{Ir}$) states to be projections of the occupied electronic states
onto the impurity states  of corresponding symmetry
centered on the substitutional atomic site:
\begin{equation}
N_{s,p,d}^{Ir}=\int_{-\infty}^{E_{f}} \rho_{s,p,d}^{Ir}(\epsilon) d\epsilon.
\label{charge}
\end{equation}
$E_{f}$ is the Fermi level. This projection is {\it not} simply onto the atomic states of the Ir atom but
onto the Wannier states centered at the Ir sites and extended in the whole supercell.
In a metal the net charge of the impurity must vanish so that
the  total occupation number of such extended orbitals,
is equal to the number of atomic valence electrons,
\begin{equation}
N_{s}^{Ir}+N_{p}^{Ir} + N_{d}^{Ir} = 9.
\label{charge}
\end{equation}
This constraint is respected by our density functional theory (DFT) calculation (see later).

According the Friedel sum rule, the phase shifts $\delta$ can be calculated by
the occupation numbers of the corresponding orbitals as \cite{Niimi-CuBi,Langreth}
\begin{equation}
\begin{split}
\delta_{1} &= \delta^{Ir}_{p} -\delta^{Cu}_{p} = \frac{\pi(N_{p}^{Ir}-N_{p}^{Cu})}{6}, \\
\delta_{2}^{+} &= \delta^{Ir}_{d+} -\delta^{Cu}_{d+} =\frac{\pi(N_{d+}^{Ir}-N_{d+}^{Cu})}{6}, \\
\delta_{2}^{-} &= \delta^{Ir}_{d-} -\delta^{Cu}_{d-}=\frac{\pi(N_{d-}^{Ir}-N_{d-}^{Cu})}{4}.
\end{split}
\label{phaseshifts}
\end{equation}

When the correlation $U$ is included into the $5d$ states of the Ir impurity,
the decreased $N_{d}^{Ir}$ is accompanied by increased $N_{s}^{Ir}$ and $N_{p}^{Ir}$, according to Eq. (\ref{charge}).
By Eq. (\ref{phaseshifts}), a negative  $\delta_{1}$ ($\sim N_{p}^{Ir} - N_{p}^{Cu}$) may change sign and
become positive, due to the increased $N_{p}^{Ir}$.
This will be confirmed by the calculations in the following sections.

\emph{Spin Hall Angle due to impurities of $d$ orbitals}--- For an electron
scattered by a potential with the SOI,
the amplitudes of the scattered wave are given by \cite{Bethe-Jackiw,Gu-CuBi}
\begin{equation}
\begin{split}
f_{\uparrow}(\theta) &= f_1(\theta) |\uparrow\rangle + e^{i\varphi}f_2(\theta)|\downarrow\rangle, \\
f_{\downarrow}(\theta) &= f_1(\theta) |\downarrow\rangle - e^{-i\varphi}f_2(\theta)|\uparrow\rangle,
\end{split}
\label{f1}
\end{equation}
for incoming spin-up and spin-down electrons. $f_1(\theta)$ and $f_2(\theta)$ represent the spin non-flip and spin flip scattering amplitudes, respectively.
$\theta$ and $\varphi$ are the polar and azimuthal angles of the scattered wave vector.
The scattering amplitudes can be expressed in terms of  the phase shifts $\delta_{l}$ of the orbitals $l$ as
\begin{equation}
\begin{split}
f_1(\theta) &= \sum_{l}\frac{P_{l}(\cos\theta)}{2ik}[(l+1)(e^{2i\delta_{l}^{+}}-1)+l(e^{2i\delta_{l}^{-}}-1)], \\
f_2(\theta) &= \sum_{l}\frac{-\sin \theta}{2ik}(e^{2i\delta_{l}^{+}}-e^{2i\delta_{l}^{-}})\frac{d}{d\cos\theta}P_{l}(\cos\theta),
\end{split}
\label{f2}
\end{equation}
with the two phase shifts $\delta_{l}^{\pm}=\delta_{l\pm \frac{1}{2}}$ differing because of spin-orbit terms;
without them $\delta_{l}^{+}=\delta_{l}^{-}=\delta_{l}$.

The spin-independent part $I(\theta)$ and skewness $S(\theta)$ of the scattering cross section are represented by
\begin{equation}
\begin{split}
I(\theta) &= |f_1(\theta)|^2 + |f_2(\theta)|^2, \\
S(\theta) &= \frac{2Im\left[f^{\ast}_1(\theta)f_2(\theta)\right]}{|f_1(\theta)|^2 + |f_2(\theta)|^2}.
\end{split}
\label{I-S}
\end{equation}
And the transport skewness $\gamma_k$ is defined as \cite{Engel}
\begin{equation}
\gamma_k=\frac{\int d\Omega I(\theta)S(\theta)\sin\theta}{\int d\Omega I(\theta)(1-\cos\theta)}.
\label{skewness}
\end{equation}

The SHA $\alpha$ can be defined either by conductivity $\sigma$ or by resistivity $\rho$ as \cite{Gu-CuBi,Engel}
\begin{equation}
\alpha(\sigma)=\sigma^{(+)}_{yx}/\sigma^{(+)}_{xx}=\frac{\gamma_{k_F}}{2},
\label{SHA1}
\end{equation}
\begin{equation}
\alpha(\rho ) =\rho^{(+)}_{yx}/\rho^{(+)}_{xx}= -\alpha (\sigma).
\label{SHA2}
\end{equation}
We note that it was $\alpha(\rho)$ that was measured in the experiment of Ir doped Cu \cite{Niimi-CuIr},
while  $\alpha(\sigma)$ was calculated in the previous theory \cite{Fedorov}.
Using  consistent definitions, the SHA in Ref. \cite{Fedorov} is  opposite in sign
to the experimental value of Ref. \cite{Niimi-CuIr}.

For the nonmagnetic impurity Ir as an extrinsic scatterer in Cu,
the skew scattering arises from
the interference between the antisymmetric scattering of $l$=2 and
the symmetric scattering of $l$=1 channels \cite{Fert}.
It is assumed that only the $d$-wave scattering is the resonant channel with appreciable SOI,
while the $p$-wave is taken spin independent.
Then substituting Eq. (\ref{f2}) with parameters of $\delta_{1}$,
$\delta_{2}^{+}$ and $\delta_{2}^{-}$ into Eq. (\ref{SHA2}),
the SHA is obtained as
\begin{equation}
\alpha(\rho ) = -\frac{6\sin\delta_1\left[\sin(\delta^{-}_{2}-\delta_1)\sin\delta^{-}_{2}-\sin(\delta^{+}_{2}-\delta_1)\sin\delta^{+}_{2}\right]}
{5\left(3\sin^2\delta^{+}_{2}+2\sin^2\delta^{-}_{2}\right)}.
\label{SHA4}
\end{equation}
We note that a factor $1/2$ was missing from the equation for the SHA due to skew scattering of $d$ orbitals in
Refs. \cite{Guo,Gu-AuFe,Gu-AuPt}, and the quoted numerical results for that angle should have been divided by 2.

\emph{LDA+SOI results.}---  We now calculate the phase shift parameters
$\delta_1$, $\delta_{2}^{+}$ and $\delta_{2}^{-}$ for the Ir impurity doped in Cu host based on DFT
with the local density approximation (LDA) plus SOI.
We use the code of Quantum Espresso (QE) \cite {QE} with Hubbard $U$= 0.
A supercell of Cu$_{26}$Ir was used to calculate the occupation numbers of the orbitals around the Ir impurity,
while the primitive cell of a single Cu atom gave the occupation numbers of the orbitals of the Cu host.
The cutoff energy is 50 Ry for ultrasoft pseudopotentials with the Perdew-Burke-Ernzerhof (PBE)
type of exchange-correlation functionals \cite{PBE},
and the energy convergence limit is 10$^{-8}$ Ry,
with a $k$ lattice of $8\times8\times8$.

From these calculations for  $U$=0,
the occupation numbers, phase shifts and SHA are listed in Table \ref{data}.
The total occupation number of Ir with $5d$, $6s$ and $6p$ states was
$N_{s}^{Ir}+N_{p}^{Ir} + N_{d}^{Ir}=9.0$,
in agreement with Eq. (\ref{charge}).
The negative sign of the obtained SHA is opposite to the positive value of +2.1\% in experiment \cite{Niimi-CuIr}.
However, the small value of $\delta_{1}$ is consistent with previous estimation of $|\delta_{1}|\simeq 0.1$ \cite{Fert,Fert-Levy}.
Motivated by the idea in Fig. \ref{F-schematic}, we will check the effect of local electron correlation $U$ in the following section.

\begin{table}
\begin{tabular}{c|ccc|ccc|c}
\hline
U (eV) & $N_{d}^{Ir}$ & $\delta_{2}^{+}$ & $\delta_{2}^{-}$ & $N_{s}^{Ir}$ & $N_{p}^{Ir}$ & $\delta_{1}$ & SHA \\
\hline\hline
{\bf 0  } & {\bf 7.82} &{\bf  -0.73} &{\bf  -0.38} &{\bf  0.32} & {\bf 0.86 }&{\bf  -0.05 } &{\bf  -1.1\%  }\\
\hline
\hline
0.1   & 7.50 & -0.76 & -0.57 & 0.41 & 1.09 & 0.07 & +0.8\% \\
\hline
0.2 & 7.20 & -0.89 & -0.61 & 0.49 & 1.31 & 0.19 & +2.5\% \\
\hline
0.3 & 6.82 & -1.06 & -0.65 & 0.59 & 1.59 & 0.33 & +4.6\% \\
\hline
0.4 & 6.38 & -1.26 & -0.70 & 0.70 & 1.92 & 0.50 & +5.4\% \\
\hline
0.5 & 5.96 & -1.45 & -0.75 & 0.82 & 2.22 & 0.66 & +3.4\% \\
\hline
\end{tabular}
\caption{
Occupation numbers ($N$), phase shifts ($\delta$), and calculated SHA for an Ir impurity in a Cu host,
with  local electron correlation $U$ on the $5d$ orbitals of Ir varying from 0 to 0.5 eV.
For  $U$=0  (first line, in boldface),
the occupation numbers are calculated with LDA+SOI by DFT,
which also gave $N_{p}^{Cu}$=0.96 and $N_{d}^{Cu}$=9.68 for the Cu host.
For $U>$0,
$N_{d}^{Ir}$, $\delta_{2}^{+}$ and $\delta_{2}^{-}$ are calculated by the QMC method.
$N_{s}^{Ir}$ and $N_{p}^{Ir}$ are then estimated from Eq. (\ref{charge}), keeping the ratio of $N_{p}^{Ir}/N_{s}^{Ir}$ as for $U$=0.
For each value of $U$ the phase shift $\delta_{1}$ follows from the Friedel sum rule, Eq. (\ref{phaseshifts}) and
the SHA from Eq. (\ref{SHA4}) in the text.}
\label{data}
\end{table}

\emph{Quantum Monte Carlo results.}---
Due to the experimental observation that
the SHA of CuIr is independent of the concentration of Ir impurities \cite{Niimi-CuIr},
we employ a single-impurity multi-orbital Anderson model with SOI
\cite{Anderson, Gu-AuFe, Gu-AuPt}
\begin{eqnarray}
  H&=&\sum_{\textbf{k},\alpha,\sigma}\epsilon_{\textbf{k}\alpha}
  c^{\dag}_{\textbf{k}\alpha\sigma}c_{\textbf{k}\alpha\sigma}
   +\sum_{\textbf{k},\alpha,\xi,\sigma}(V_{\xi\textbf{k}\alpha}
    d^{\dag}_{\xi\sigma} c_{\textbf{k}\alpha\sigma} + H.c.) \notag\\
  &+& \sum_{\xi,\sigma}\epsilon_{\xi}n_{\xi\sigma}
   +U\sum_{\xi}n_{\xi\uparrow}n_{\xi\downarrow}
   + \frac{U^{\prime}}{2}\sum_{\xi\neq\xi',\sigma,\sigma^{\prime}}
     n_{\xi\sigma}n_{\xi'\sigma^{\prime}} \notag\\
   &-& \frac{J}{2}\sum_{\xi\neq\xi',\sigma}n_{\xi\sigma}n_{\xi'\sigma}
  +\frac{\lambda}{2}\sum_{\xi,\sigma}d^{\dagger}_{\xi\sigma}
   (\ell)^{z}_{\xi\xi}(\sigma)^{z}_{\sigma\sigma}d_{\xi\sigma},
\label{andersonmodel}
\end{eqnarray}
where $\epsilon_{\textbf{k}\alpha}$ is the energy band $\alpha$ of host Cu,
$\epsilon_{\xi}$ is the energy level of the 5d orbital $\xi$ of impurity Ir,
$V_{\xi\textbf{k}\alpha}$ is the hybridization between the 5d orbital $\xi$ of Ir and the band $\alpha$ of Cu,
and $\lambda$ is the strength of SOI.
$U$ ($U^{\prime}$) is the on-site Coulomb repulsion within (between) the $5d$ orbitals of Ir,
and $J$ is Hund coupling between the $5d$ orbitals of Ir.
The $\epsilon_{\textbf{k}\alpha}$, $\epsilon_{\xi}$ and $V_{\xi\textbf{k}\alpha}$ can be
obtained from the codes of QE \cite{QE} and Wannier90 \cite{wannier90}, respectively \cite{GuCon}.

We apply the quantum Monte Carlo (QMC) method \cite{Hirsch-Fye,QMCreview,GuCon},
which can correctly include the local electron correlations, to calculate the occupation numbers
$n_{\xi}$ of $5d$ orbitals of Ir.
The parameters of correlation and SOI of Ir are given to be $U$=0.5 eV and $\lambda$=0.5 eV \cite{UIr,UIr2}.
The relations of $U=U^{\prime}+2J$ and $J/U=0.3$ \cite{Maekawa}
give $J$=0.15 eV and $U^{\prime}$=0.2 eV.

In order to allow convergence down to room temperature in a tractable calculation, of the five $5d$ orbitals of Ir  only the
three $t_{2g}$ orbitals were retained, and only the diagonal component of the SOI is conserved.
It is  reasonable to neglect the $e_{g}$ orbitals  because there are  no spin-orbit matrix elements.
It is convenient to transform the three $t_{2g}$ orbitals into  $t_{-1}$, $t_{0}$ and $t_{1}$:
$t_{-1} \equiv -\frac{1}{\sqrt{2}}(xz-iyz)=-Y_{2,-1}$,
$t_{0} \equiv -ixy=-\frac{1}{\sqrt{2}}(Y_{2,2}-Y_{2,-2})$
and $ t_{1} \equiv -\frac{1}{\sqrt{2}}(xz+iyz)=Y_{2,1},$
which are expressed by spherical harmonics with the orbital angular momentum $\ell^{z}=$-1, 0 and 1 of $l$=2, respectively.
The QMC calculation is performed with more than 10$^{5}$ Monte Carlo sweeps,
the Matsubara time step $\Delta \tau$=0.25, and temperature T=360 K.

The QMC calculation \cite{Gu-AuFe,Gu-AuPt} can give the occupation number $\langle n_{-1} \rangle$, $\langle n_{0} \rangle$
and $\langle n_{1} \rangle$ for each $t_{2g}$ orbital.
The average value of the $z$ component of spin-orbit correlation function
$\langle \ell^{z}\sigma^{z} \rangle \equiv \langle -n_{-1\uparrow}+n_{-1\downarrow}+n_{1\uparrow}-n_{1\downarrow} \rangle$,
can also be obtained.
The occupation numbers of parallel and antiparallel states can be written as
$n_{P} = \langle n_{-1\downarrow} \rangle+\langle n_{1\uparrow} \rangle$,
$n_{AP} = \langle n_{-1\uparrow} \rangle+\langle n_{1\downarrow} \rangle$,
which can be obtained from the relations of
$\langle \ell^{z}\sigma^{z} \rangle =n_{P}-n_{AP}$ and
$\langle n_{1}\rangle+\langle n_{-1}\rangle = n_{P}+n_{AP}$.
These occupation numbers are related to the the phase shift of
the parallel (antiparallel) state of the $t_{2g}$ orbitals of Ir as
$\delta_{P} = \pi n_{P}/2$ ($\delta_{AP} = \pi n_{AP}/2$).
We assume that $\delta_{P}$ (resp. $\delta_{AP}$) from the simplified model
can be taken as the phase shift $\delta_{d+}^{Ir}$ ($\delta_{d-}^{Ir}$)
of the full spin-orbit split $5d$ orbital of Ir.

With correlation $U$=0.5 eV on the $5d$ states of Ir impurity,
the QMC calculation gives $\delta_{d+}^{Ir}=1.59$ and $\delta_{d-}^{Ir}=2.30$.
The occupation numbers can be calculated as $N_{d+}^{Ir}=6\delta^{Ir}_{d+}/\pi = 3.03$ and
$N_{d-}^{Ir}=4\delta^{Ir}_{d-}/\pi = 2.93$,
and the phase shifts of $\delta_{2}^{+}$ and $\delta_{2}^{-}$ then obtained by Eq. (\ref{phaseshifts})
to be -1.45 and -0.75, respectively, as listed in the case of $U$=0.5 eV in Table \ref{data}.
Comparing the cases with $U$ of 0.0 and 0.5 eV in Table \ref{data},
it is clear that the splitting of the phase shifts,
$|\delta_{2}^{+}-\delta_{2}^{-}|$, is enlarged
from 0.35 to 0.70 by the correlation $U$=0.5 eV.
Moreover, the occupation number of $5d$ states is
$N_{d}^{Ir} = N_{d+}^{Ir} + N_{d-}^{Ir} = 5.96 $, which is smaller than
the $N_{d}^{Ir} = 7.82 $ without $U$, in agreement with the schema in Fig. \ref{F-schematic}.

The decreased electron number in $5d$ states, which is 1.86,
will be transferred to the $6s$ and $6p$ states of Ir impurity as discussed by Eq.(\ref{charge}).
We do not know precisely how many electrons are transferred  to the $6s$ states,
and how many  to the $6p$ states. As an approximate estimation \cite{footnote2},
we fix the ratio $N_{p}^{Ir}/N_{s}^{Ir}$=2.7
which is obtained by LDA with $U$=0.
This  gives the increased $N_{p}^{Ir}$= 2.22, as shown in the case of $U$=0.5 eV in Table \ref{data}.
Since the occupation numbers for pure Cu ($N_{p}^{Cu}$, $N_{d}^{Cu}$) do not change,
the data of $N_{p}^{Cu}$=0.96 and $N_{d}^{Cu}$=9.68 obtained from the $U$=0 case
are employed for calculating $\delta_{1}$ and SHA for all positive $U$ in Table \ref{data}.
Finally,  positive $\delta_{1}$ of 0.66 and  SHA of +3.4\% follow from Eqs. (\ref{phaseshifts}) and (\ref{SHA4}),
respectively, for $U$=0.5 eV.
The SHA agrees in sign, but overestimates the experimental of +2.1\% in magnitude \cite{Niimi-CuIr}.
Our theory indicates, then, that the correlation $U$ on the $5d$ states of Ir impurity
is the crucial factor to give a positive phase shift $\delta_{1}$ and a positive SHA as in experiment.

\begin{figure}[tbp]
\includegraphics[width = 6.5 cm]{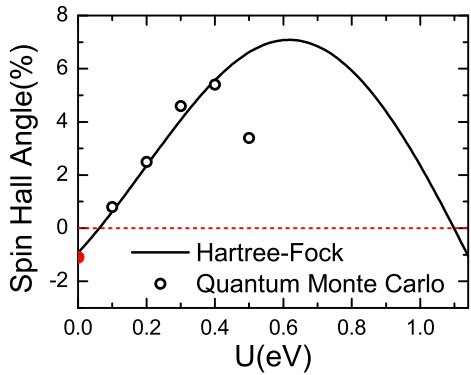}
\caption{(color online). Calculated SHA by LDA under $U$=0 (red dot), by QMC under $U>$0 (black circles) from Table \ref{data},
and by the Hartree-Fock approximation for spin-orbit-split $5d$ orbitals (black line).}
\label{HF}
\end{figure}

To show the change of SHA between cases of $U$=0 and $U$=0.5 eV, intermediate values of  $U$ were calculated. Keeping the condition in Eq.(\ref{charge}) and fixing the ratio $N_{p}^{Ir}/N_{s}^{Ir}$=2.7,
the resulting SHA by Eq. (\ref{SHA4}) ranges from -1.1\% to +5.4\%, as listed in Table \ref{data} ad plotted in  Fig. \ref{HF}.
An $U$ as small as 0.1 eV would be enough to change the sign of SHA.
As for the discrepancy in magnitude compared to experiment
for the more realistic value of $U$=0.5 eV,
the energies of the the $5d$ states of Ir without correlation were determined by LDA,
which tends to overestimate the level of the states under the Fermi level.
The level of the $5d$ states of Ir is then too close to the Fermi level.
This then overestimates the  decrease of $N_{d}^{Ir}$
for a given correlation $U$. Errors in the predicted magnitude of the SHA may  also come from
the simplification from  five $5d$ orbitals to three $t_{2g}$ orbitals of Ir
in the Hamiltonian in Eq. (\ref{andersonmodel}),
as well as  uncertainty in the parameters of $U$ and $N_{p}^{Ir}/N_{s}^{Ir}$.
Accurate $U$ and $N_{p}^{Ir}$ might be measured by the X-ray spectroscopy \cite{xray1}.

For comparison, we also show  in Fig. \ref{HF}
the SHA  using  a  Hartree-Fock calculation of the occupation numbers of the full set of spin-orbit split d orbitals,
using the same values of $U^{\prime}$ and $J$
as in the QMC and obtaining the width $\Delta$ and levels for $E_{0,d^{+}}$ and $E_{0,d^{-}}$ from LDA,
but neglecting crystal field splitting:
\begin{equation}
\begin{split}
E_{d\pm}
=&E_{0,d\pm}+U(\frac{3}{5}n_{d+}+\frac{2}{5}n_{d-})+U^{\prime}(\frac{24}{5}n_{d+}+\frac{16}{5}n_{d-})\\
&-J(\frac{12}{5}n_{d+}+\frac{8}{5}n_{d-}),
\end{split}
\label{hfeq}
\end{equation}
We note that for the entire range of $U$ shown in Fig. \ref{HF},
the non-magnetic solution is stable following Ref. \cite{Anderson}.
This shows that the change of sign and non-monotonic behavior of the SHA as a function of U are {\it not}
a result of the projection onto $t_{2g}$ states but more general.

\emph{Discussion}---
Since the sign of SHA is sensitive to the sign of $\delta_{1}$ [Eq. (\ref{SHA4})]
and the corresponding small change of $N^{Ir}_{p}$ [Eq. (\ref{phaseshifts})],
the sign of SHA might be controlled as long as the occupation number of the $6p$ states of the impurity were properly manipulated.
For instance, a laser pulse \cite{laser1} can decrease the occupation number of the impurity by excitation.
An improved combination of noble metal hosts and $5d$ metal impurities
with a long lifetime of the excited states may be imagined as a means to control the sign of the SHA.

In summary,
we reconsider the theory of spin Hall effect in CuIr alloys by the QMC method,
where the local Coulomb correlation $U$ in $5d$ states of Ir impurities is included.
Taking $U$ to be 0.5 eV, we obtain a positive SHA, consistent with experiment,
in contrast to the negative angle predicted without correlation $U$.
Our result reveals the key physics determining the spin Hall effect in CuIr alloys,
explaining the positive sign of the SHA in experiment.
This may also open up a way to control the sign of the spin Hall effect by manipulating the occupation number
of the impurities.

We are very grateful to A. Fert, P. Levy, M. Chshiev, H. X. Yang, Y. Otani, and Y. Niimi for many stimulating discussions.
T. Ziman would like to thank the KITP, UCSB for hospitality during the Spintronics Program 2013,
with support in part by the National Science Foundation under Grant No. NSF PHY11-25915.
The work has been supported by a REIMEI project of JAEA.



\begin{thebibliography}{99}
\bibitem{footnote2} Due to the fact that the s and p waves of Ir are quite extended compared to the d waves,
we assume the same ratio of $N_{p}^{Ir}/N_{s}^{Ir}$ in the model with finite U on $5d$ around 0.5 eV.
The sign of SHA with U=0.5 eV is not sensitive to the ratio,
which is confirmed by varying the ratio around 2.7 by 50\%,
from 1.4 to 4.1. For the ratio of 1.4, it gives $N_{p}^{Ir}$=1.78 and SHA=+4.1\%;
for the ratio of 4.1, it gives $N_{p}^{Ir}$=2.45 and SHA=+2.2\%.
\end{thebibliography}
\end{document}